\documentclass{article}

\usepackage{arxiv}

\usepackage[utf8]{inputenc} 
\usepackage[T1]{fontenc} 
\usepackage{hyperref}
\usepackage{url}    
\usepackage{booktabs} 
\usepackage{amsfonts} 
\usepackage{nicefrac}  
\usepackage{microtype}
\usepackage{lipsum}      
\usepackage{graphicx}
\usepackage[numbers,sort&compress]{natbib}
\setcitestyle{numbers}
\usepackage{amsmath}
\usepackage{cleveref}
\usepackage{doi}
\usepackage{setspace} 
\usepackage{array}
\title{High-Throughput Detection of Risk Factors to Sudden Cardiac Arrest in Youth Athletes: A Smartwatch-Based Screening Platform}

\newif\ifuniqueAffiliation
\uniqueAffiliationtrue

\ifuniqueAffiliation
\author{ \href{https://orcid.org/0009-0002-1224-9076}{\includegraphics[scale=0.06]{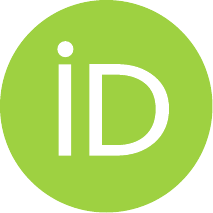}\hspace{1mm}Evan K. Xiang } \thanks{Authors are listed in order of contribution: Evan Xiang, Thomas Wang, Vivan Poddar. Contact the corresponding author at 26xiange@shadysideacademy.org.}\\
	Shady Side Academy\\
	Pittsburgh, PA 15238 \\
	\texttt{26xiange@shadysideacademy.org} 
	\And
	\href{https://orcid.org/0009-0009-3278-8625}{\includegraphics[scale=0.06]{orcid.pdf}\hspace{1mm}Thomas J. Wang} \\
	Shady Side Academy\\
	Pittsburgh, PA 15238\\
	\texttt{26wangt@shadysideacademy.org} 
        \And 
        \href{https://orcid.org/0009-0002-3425-495X}{\includegraphics[scale=0.06]{orcid.pdf}\hspace{1mm}Vivan N. Poddar} \\
	Shady Side Academy\\
	Pittsburgh, PA 15238\\
	\texttt{26poddarv@shadysideacademy.org} 
} 

\date{}

\renewcommand{\headeright}{Technical Report}
\renewcommand{\undertitle}{Technical Report}

\hypersetup{
pdftitle={High-Throughput Detection of Risk Factors to Sudden Cardiac Arrest in Youth Athletes: A Smartwatch-based Screening Platform},
pdfsubject={Cardiovascular Medicine},
pdfauthor={Evan K. Xiang, Thomas J. Wang, Vivan N. Poddar},
pdfkeywords={ECG, Diagnostics, Smartwatch, AI},
}

\begin{document}
\maketitle

\begin{abstract}
	Sudden Cardiac Arrest (SCA) is the leading cause of death among athletes of all age levels worldwide. Current prescreening methods for cardiac risk factors are largely ineffective, and implementing the International Olympic Committee recommendation for 12-lead ECG screening remains prohibitively expensive. To address these challenges, a preliminary comprehensive screening system (CSS) was developed to efficiently and economically screen large populations for risk factors to SCA. A protocol was established to measure a 4-lead ECG using an Apple Watch. Additionally, two key advances were introduced and validated: 1) A decomposition regression model to upscale 4-lead data to 12 leads, reducing ECG cost and usage complexity. 2) A deep learning model, the Transformer Auto-Encoder System (TAES), was designed to extract spatial and temporal features from the data for beat-based classification. TAES demonstrated an average sensitivity of 95.3\% and specificity of 99.1\% respectively in the testing dataset, outperforming human physicians in the same dataset (Se: 94\%, Sp: 93\%). Human subject trials (n = 30) validated the smartwatch protocol, with Bland-Altman analysis showing no statistical difference between the smartwatch vs. ECG protocol. Further validation of the complete CSS on a 20-subject cohort (10 affected, 10 controls) did not result in any misidentifications. This paper presents a mass screening system with the potential to achieve superior accuracy in high-throughput cardiac pre-participation evaluation compared to the clinical gold standard.
\end{abstract}

\section{Introduction}
\begin{figure} [!hbtp]
    \centering
    \includegraphics[width=0.75\linewidth]{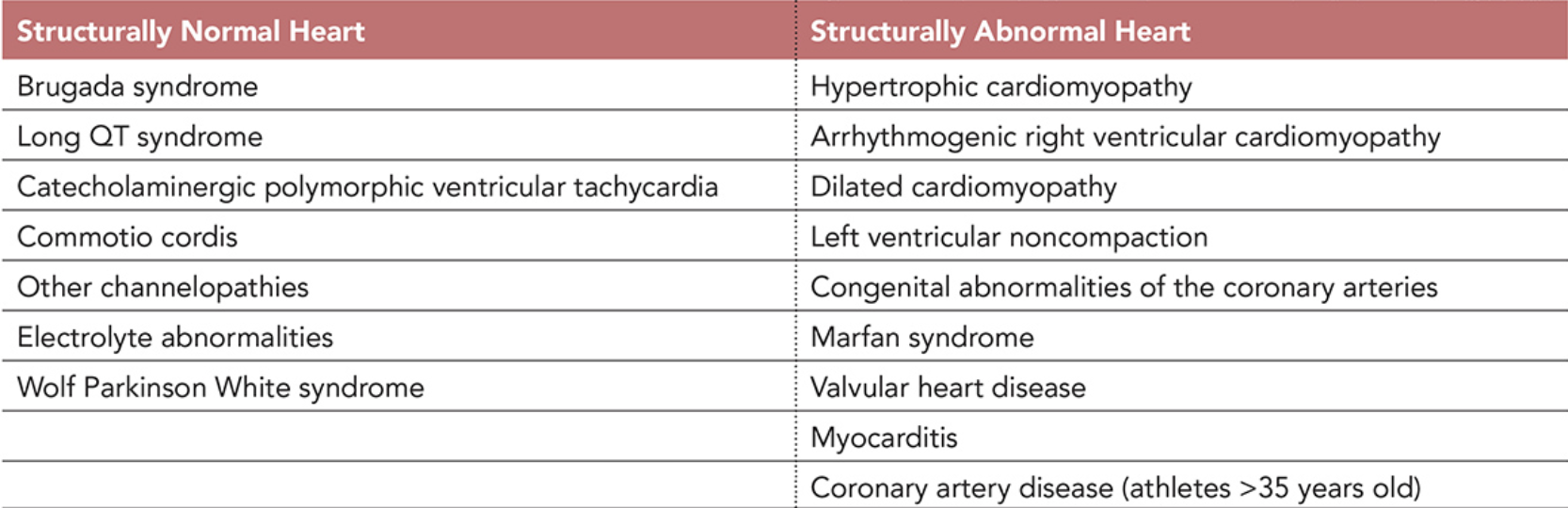}
    \caption{Common Causes of Sudden Cardiac Arrest within the Athlete Population \cite{mejialopez_2019_focus}}
    \label{fig:scacause}
\end{figure}

The National Institute of Health defines Sudden Cardiac Arrest (SCA) as a moment when the heart is not beating sufficiently to maintain perfusion due to the heart's electrical or mechanical failure \cite{Yow2022-cc}. SCA is the leading cause of death among youth athletes --- a focus group that has a heightened risk of SCA --- with 1 in 16,000 young athletes and 1 in 5200 athletes at the elite level afflicted yearly \cite{Yow2022-cc, Ha2022}. For youth athletes, the primary cause of SCA is hypertrophic cardiomyopathy (HCM) in the U.S. and arrhythmogenic right ventricular cardiomyopathy (ARVC) in Europe. SCA may also result from coronary artery disease, Long QT Syndrome, Myocarditis, Wolff-Parkinson-White syndrome, and dilated cardiomyopathy \cite{Yow2022-cc, Ha2022, maron_2003_sudden, maron_2009_sudden}. Figure \ref{fig:scacause} provides a comprehensive list of significant predictors of SCA \cite{mejialopez_2019_focus}. While these disorders do not always lead to instances of SCA, they present a substantial increase in the chance of SCA events, which is further amplified by the innate risk of sports participation \cite{Saul2010,adel_2022_risk, guasch_2023_something, a2009_the}. Concerningly, the current 14-point questionnaire pre-participation evaluation (PPE) recommended by the American Heart Association (AHA) is ineffective at detecting risk factors with poor sensitivity and specificity of 18.8\% and 68.0\% respectively \cite{Williams2019,petek_2020_current}. Numerous articles, reports, and organizations have indicated a strong and definitive call for the establishment of alternatives to the 14-point PPE \cite{Williams2019, Harmon2015, Albiski2022}. The 12-lead ECG has been promoted by the International Olympic Committee and the European Society of Cardiology as the gold standard for PPE \cite{a2009_the, Maron2014}, which is supported by a meta-analysis by Harmon et. al\cite{Harmon2015}. Harmon found that 12-lead ECG screening exhibits an average sensitivity of 94\% and specificity of 93\%, significantly outperforming the 14-point PPE. 

The ECG is a non-invasive test that measures the electrical activity of the heart, most commonly using 12 electrode leads attached to various parts of the body. In regions of Italy and Switzerland, the 12-lead ECG has been used as a part of PPE mass screening for nearly a decade on adolescents, reducing cases of SCA by 89\% \cite{Albiski2022}. However, the American College of Cardiology does not recommend ECG mass screening due to two key factors:
\begin{enumerate}
    \item High False Positive Rate (FPR): The 7\% FPR of the 12-lead ECG towards detecting SCA risk factors is exceedingly high for implementation in mass-screening \cite{Harmon2015, Wheeler2010}. Economically, this is not feasible, as the cost of follow-up appointments for more invasive confirmative procedures is substantial \cite{OConnor2010}. \citeauthor{OConnor2010} also finds that 98.9\% of the cost of a nationwide ECG screening program would come from follow-up screening on false-positive cases, indicating that the FPR of ECG must be decreased to meet economic standards. 
	\item Healthcare Infrastructure: In the U.S., the availability of qualified medical professionals is insufficient for ECG mass screening \cite{Maron2014}. In addition, conventional ECG analysis is already time-consuming, labor-intensive, and inaccurate \cite{Mant2007}. The complex composition of ECG signals presents a challenge for even the most experienced general practitioners to analyze, achieving 80\% sensitivity and 92\% specificity in the diagnosis of atrial fibrillation \cite{Mant2007}. Therefore, a simpler, self-screening protocol to obtain high-fidelity ECGs would reduce strain on the U.S. healthcare infrastructure \cite{Maron2014}.
    
\end{enumerate}

Existing research has demonstrated the viability of acquiring 4-lead ECG data via Smartwatches (SW) and computationally reconstructing 12-lead ECG from a limited number of leads. \citeauthor{Touiti2023} and \citeauthor{Behzadi2020} suggested that a lower lead smartwatch-based ECGs may be optimal for mass-screening, as lead placement for the 12 lead ECG takes significant time and has high associated cost. \citeauthor{Sohn2020} concluded that a 3-lead ECG can achieve equivalent results to a 12-lead ECG using a long short-term memory (LSTM) network, resulting in a correlation coefficient of 0.95. Two years later, \citeauthor{Jain2022} proposed a method to upscale fewer leads to 12 leads using a linear interpolation method, iteratively updating lead weights in a matrix via gradient descent, achieving a mean correlation coefficient of 0.83. 

Additionally, several studies using a variety of machine learning (ML) techniques have achieved comparable or superior classification accuracy versus physicians. \citeauthor{Smigiel2021} performed signal classification on the PTB-XL ECG database, using convolutional neural networks (CNN) with/without entropy and SincNet, yielding sensitivity and specificity of around 80\%. Many studies focus on the use of CNN, recurrent neural networks (RNN), LSTM, and hybrids of these models for ECG classification \citep{Prabhakararao2022, 8688435, Oh2018, Tan2018, Pokaprakarn2022, Rahhal2016}. However, CNN is not ideal for addressing sequential ECG data and RNN faces issues with delayed information transfer and long-term dependencies, making both networks flawed when dealing with complex ECG signals \cite{Zihlmann2017, Ding2023}. One approach by \citeauthor{9171084} extracted statistical features from ECG signals and used random forest tree algorithms to achieve an accuracy of 97.45\% in classifying 11 types of arrhythmia. Using variants of the transformer architecture, \citeauthor{Wang2021}, \citeauthor{Che2021}, and \citeauthor{Hu2022} achieved high accuracy in arrhythmia detection. In 2023 \citeauthor{Ding2023} utilized a Transformer Stacked Auto-encoder model, achieving an F1 score of 99.13\% in arrhythmia detection. These studies demonstrate ML and transformers, specifically, are suited for ECG signal classification.

However, there has been no attempt to combine these findings to validate the feasibility of a comprehensive tool for high throughput screening. The versatility of existing solutions must be improved to meet economic standards. We intend to address this gap by creating a cost-effective comprehensive screening solution and establishing its preliminary validity in silico and in human subjects.

\newpage
\section{Methodology}
\begin{figure}[!h]
    \centering
    \includegraphics[width = \linewidth]{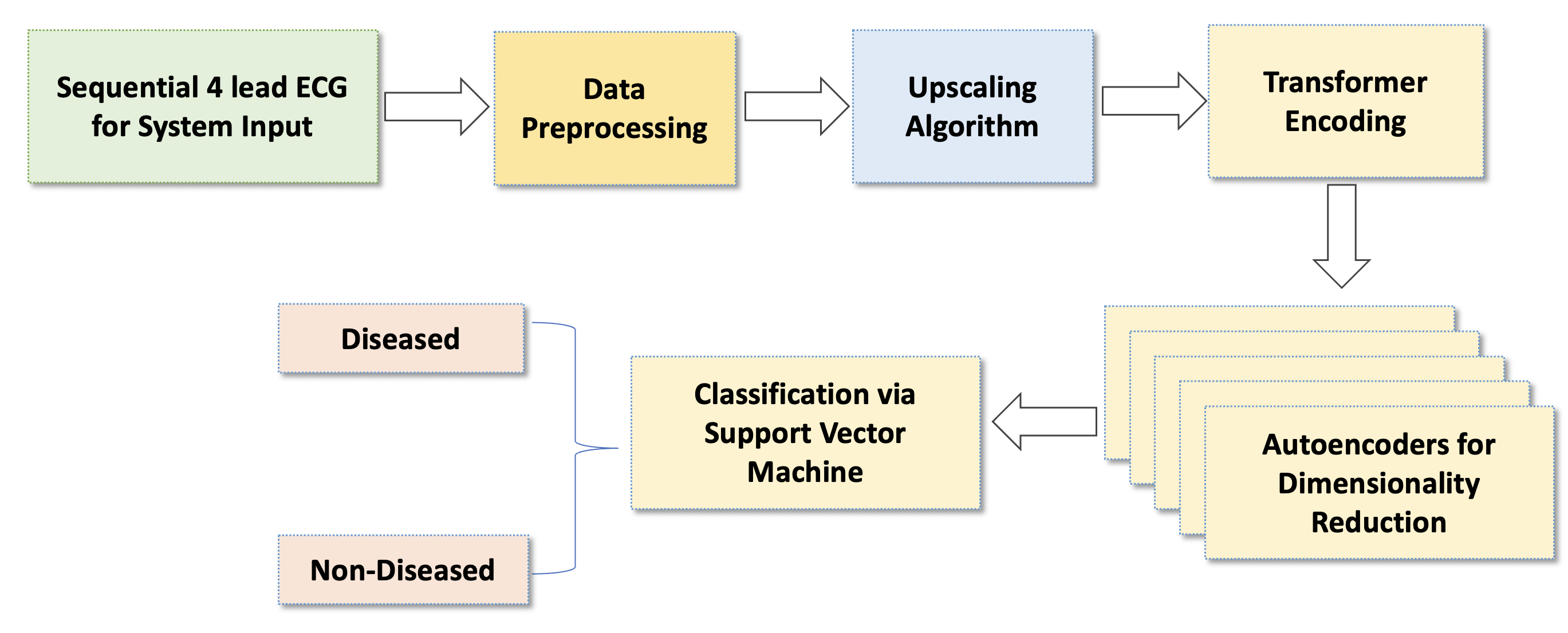}
    \caption{Brief Overview of Research Methodology. A sequential 4-lead ECG is taken. It is then preprocessed as described below, upscaled to 12 leads, and then spatial and temporal features are extracted using the TAES model. OvO SVM is used to classify the final result as the N (normal) label or the H, M, L, D, A (hypertrophic cardiomyopathy, myocarditis, long QT syndrome, dilated cardiomyopathy, arrhythmogenic left ventricular hypertrophy) labels. }
    \label{fig:overview}
\end{figure}

Figure \ref{fig:overview} briefly overviews our methodology. A protocol is developed for taking sequential 4-lead ECGs using the Apple Watch S7. The extracted data is then rescaled and resampled in accordance with typical signal processing. An upscaling algorithm using decomposition-based regression is implemented to upscale our data from 4 to 12 leads. The classification algorithm, TAES (Transformer Auto-Encoder System), classifies the ECG via extraction of a spatial and temporal feature set. Finally, to validate our watch protocol, ECG screening is conducted on 30 participants, followed by comparative ECG interpretation on a sub-cohort of 20 participants (10 afflicted, 10 controls) to determine the baseline viability of our algorithms. 

\subsection{Apple Watch S7 Trace Recording} \label{protocol}

\begin{figure} [!htbp]
    \centering
    \includegraphics[width=0.7\linewidth]{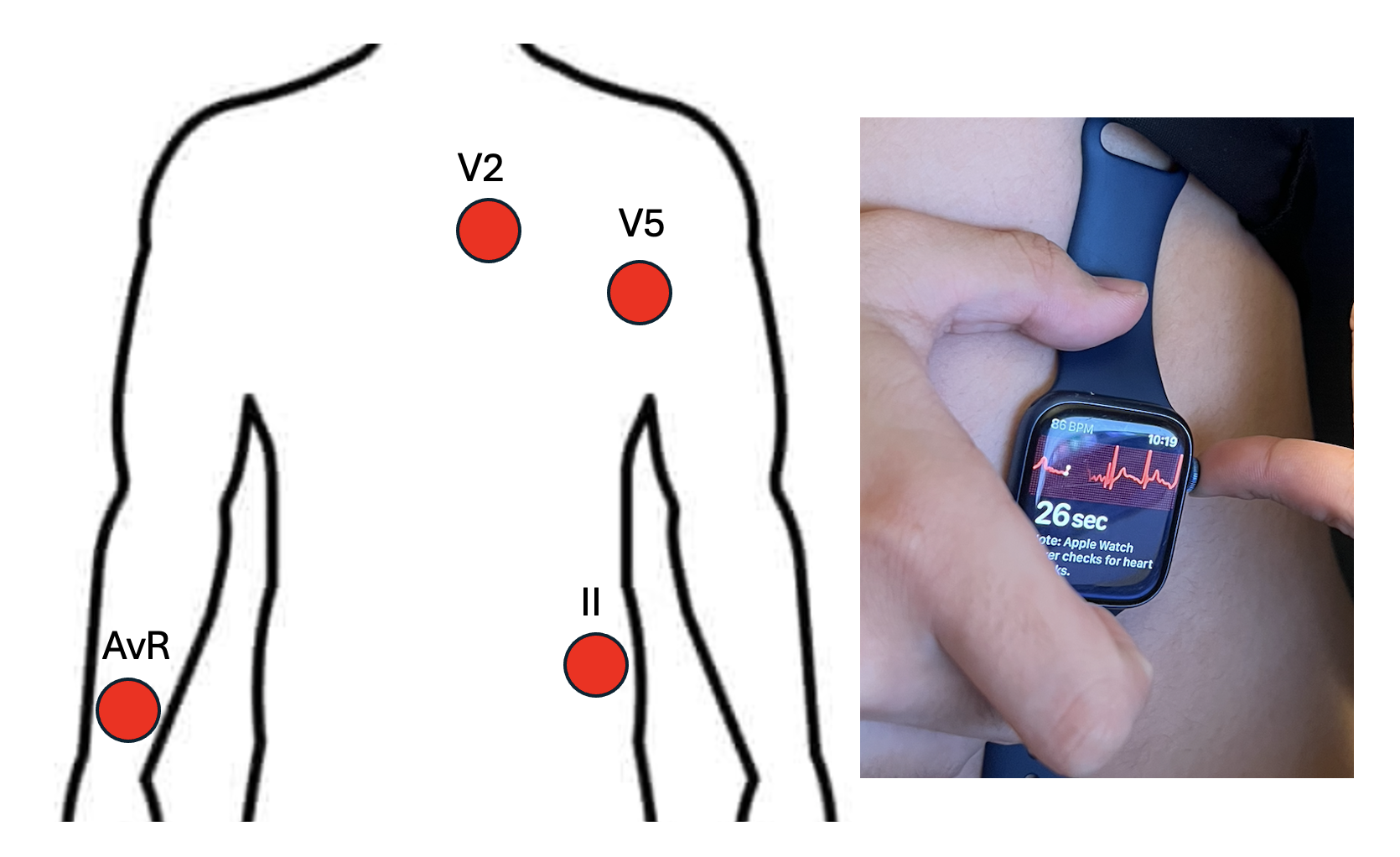}
    \caption{Description of S7 Placement in the Sequential ECG Protocol. Leads V2, V5, II, and AvR were chosen to be measured. The left depicts the typical position in which the device is held during measurement.}
    \label{fig:humanexample}
\end{figure}

It is possible to record several ECG leads via the differential positioning of a smartwatch. An Apple Watch Series 7 (MSRP: 399 USD) is utilized as the watch of choice for our screening system due to its wide commercial availability in the US. The S7 can take a one-lead ECG using the digital crown as the negative electrode and the base of the watch as the positive electrode. A protocol was designed with the S7 to take a sequential 4-lead ECG. The specific leads may be measured in the manner described in Figure \ref{fig:humanexample}. Further clarification on the designed protocol is provided below: 
\begin{itemize}
    \item \textbf{Lead II:} Lead II is measured by placing the Apple Watch on the lower left abdomen, just above the hip, with the right finger on the digital crown. Typically, Lead II is derived from the difference between the Right Arm (RA) electrode and the Left Leg (LL) electrode. In our case, we utilize the placement on the lower abdomen to simulate the LL electrode while using the right finger on the digital crown to simulate the RA electrode.
    \item \textbf{Lead AvR:} Lead AvR is measured by placing the watch on the right arm with the left finger on the digital crown. This simulates the typical AvR lead which is measured using the difference between the right arm (RA) electrode and the left arm (LA) electrode
    \item \textbf{Lead V2:} Lead V2 is measured by placing the watch at the 4th intercostal space, positioned about 2 centimeters to the left of the sternum. A typical procedure was followed, using the index finger of the right arm as the negative electrode. This positioning allows this lead to view the anterior septal wall, simulating a Wilson-type chest lead.
    \item \textbf{Lead V5:} Lead V5 is measured by placing the watch at the 5th intercostal space along the anterior axillary line. The right finger is placed on the digital crown. This is typically located about 5 inches underneath the armpit in a horizontal line with the clavicle, simulating a Wilson-type chest lead.
\end{itemize}

\subsection{Data Pre-Processing and Training}
All data was acquired from a variety of publicly available datasets on PhysioNet, ultimately consisting of 41,350 12-lead ECG of unstandardized length between 5 seconds and over 30 minutes. Data was stratified in terms of age and sex. All data was sampled at 250 Hz. Only patients with normal sinus rhythm, hypertrophic cardiomyopathy (H), dilated cardiomyopathy (D), arrhythmogenic left ventricular hypertrophy (A), myocarditis (M), or long QT syndrome (L), were included in the dataset. 

\subsubsection{Segmentation and Preparation}
Baseline normalization was performed, followed by beat-based segmentation of the ECG files. Each file was segmented to a single heartbeat just before the R peak, identified using the algorithm described in \citeauthor{Pooyan2016}. Data was spliced from 49 data points prior to the identified R peak to 50 points post-beat. This resulted in 3,236,931 files in the primary dataset. As there were multiple pieces of data per patient, all pieces of the same ECG were segmented into the same split. For example, all beats from one recording would end up in the testing set, and all beats of another may end up in the testing set. Z-score normalization is then performed to ensure equal algorithmic input. 

\subsubsection{Data Filtration}
The following steps were used on both the external training data and the watch-derived data following beat-based segmentation. 
\begin{enumerate}
    \item Butterworth bandpass filter between 2 Hz and 40 Hz to remove baseline wander, electrode wander, and excess physiological noise. 
    \item Powerline filter at 60 Hz to mitigate interference from any nearby electronic devices
    \item Resample all data to the range [-1,1] via the below equation to neutralize the effects of irregular maxima and minima:
        $$f(x)=2\times \frac{x-min(x)}{max(x)-min(x)}$$
    \item Downsample data to 100 Hz via Fast Fourier Transform to conserve computing power. This step may be mitigated with higher computing power.  
\end{enumerate}

\subsubsection{Data Split}
Data was separated into the following splits: 98\% of the data was utilized for the training data. 1\% was reserved for the validation set. 1\% was reserved for the testing set. Since a large amount of data was provided, 1\% of the data provides enough variance for us to draw conclusions from the results of the model training. 

\subsection{Training}
Training for both algorithms was conducted on a PC with the following specifications: AMD Ryzen 7 3700X 8-Core Processor at 4.20 GHz, 32GB of RAM at 3600 mHz, and NVIDIA GeForce RTX 3070 GPU. All experiments are carried out on this PC. The upscaling algorithm was trained by taking 12 lead data, splicing to 4 lead, iteratively feeding through the algorithm and minimizing the difference in mV between the synthesized and real ECG using MSE loss. The algorithm was tested on the validation set every 5 epochs to prevent overfitting. The classification algorithm was trained by cutting data from 12 lead ECG data to 4 leads, then using the upscaling weights to resynthesize the 12 lead ECG. This approach allows the algorithm to become accustomed to common biases within the upscaling algorithm. The algorithm was tested on the validation set every 10 epochs to prevent overfitting. In addition to these parameters, L1 regularization was applied globally to mitigate overfitting. Specific optimal parameters for the classification algorithm are discussed in the results section. 

\begin{figure} [!hbtp]
    \centering
    \includegraphics[width=0.65\linewidth]{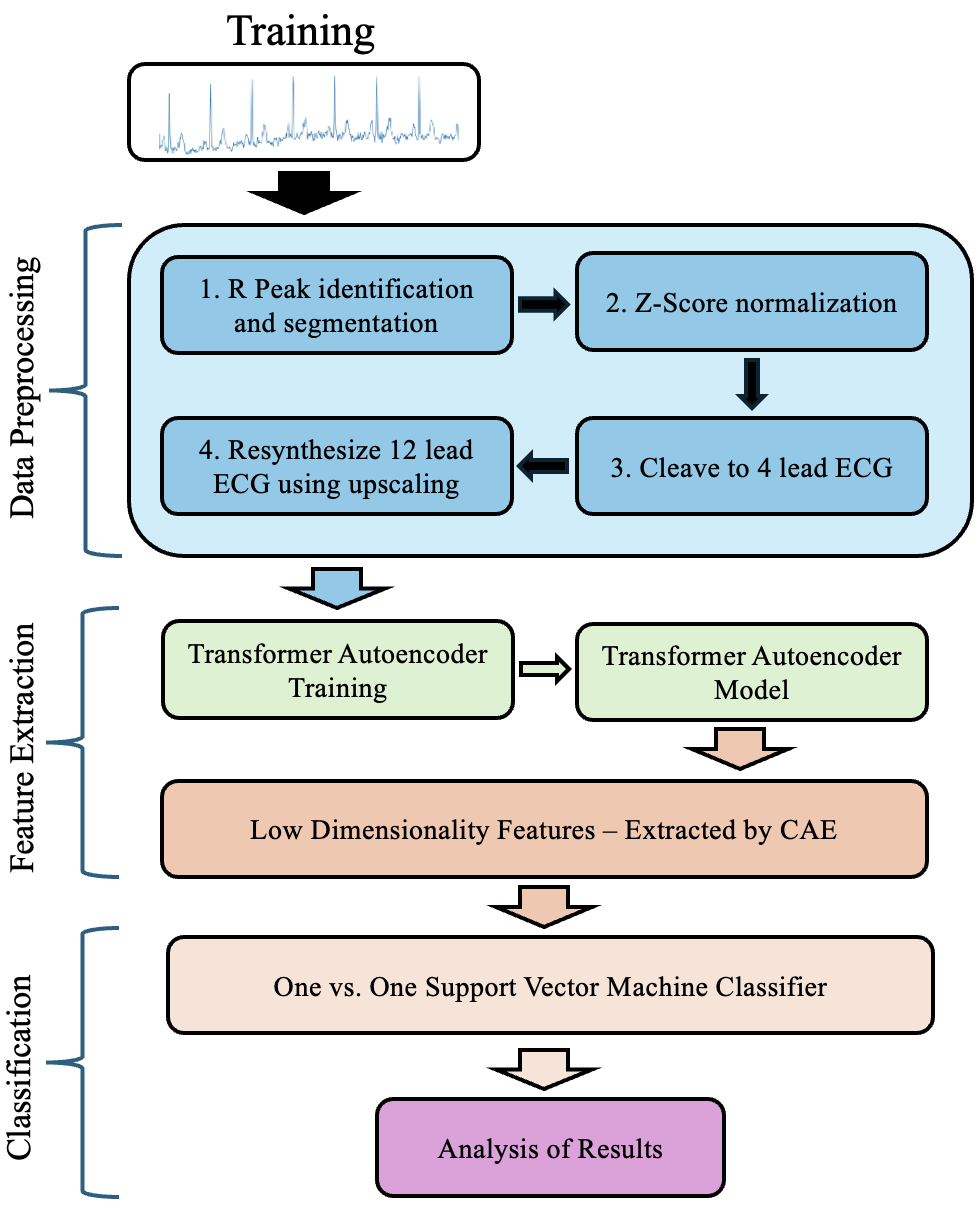}
    \caption{Overview of the Training Process. All data is fed into the preprocessing algorithm in MATLAB. Data is then retrieved and used for autoencoder model training in Python. The finalized Transformer-Autoencoder is used to retrieve low-dimensional latent state representations which are then used by a One vs. One SVM for classification. The final results are analyzed by comparing sensitivity, specificity, and F1 score.}
    \label{fig:training}
\end{figure}

\subsection{Decomposition-Based Regression - Upscaling}
The proposed algorithm uses a regression-type approach to synthesize the remaining leads from a 4-lead ECG. Certain ECG leads correspond to similar sections of the heart, thus it is predicted that a singular lead may be used to represent each major section of the heart. Since physiological data is inherently non-linear, cubic regression is used to create non-linear fits that relate the 4 chosen leads to each lead to be synthesized. This is described generally by the equation below: 
\[L = a + \sum_{i=1}^{k} \left(b_{i1} X_i + b_{i2} X_i^2 + b_{i3} X_i^3 \right) + \epsilon\]
In this equation, L is the target lead to be derived, with X representing the known leads (aVR, II, V2, V5). $b_{i}$ represents the regression coefficients that are iteratively updated. $\epsilon$ represents the error term and a represents the bias term. This equation is then expanded to all leads to be synthesized. To begin training, we instantiate a matrix with random weights as shown below:
\[ W = 
\begin{bmatrix}
0.15 & 0.82 & 0.43 \\
0.57 & 0.33 & 0.78 \\
0.61 & 0.25 & 0.14 \\
0.89 & 0.45 & 0.91 \\
0.12 & 0.67 & 0.38 \\
0.75 & 0.29 & 0.56 \\
0.48 & 0.11 & 0.64 \\
0.36 & 0.59 & 0.72 \\
\end{bmatrix} \]
The parameters of the matrix are then iteratively optimized by minimizing a loss function. Specifically, a mean square error (MSE) loss function is defined to quantify error between the predicted lead \(\hat{L}\) and actual lead \(L\). 
\[ \mathcal{L} = \frac{1}{n} \sum_{j=1}^{n} (L_j - \hat{L}_j)^2\]
When starting training, the random matrix $W$ is used as the parameter set to generate a synthesized output. The gradient of the loss function above is then calculated with respect to each parameter within the parameter set $W$. Finally, the Adam optimizer is used to perform weight updates in accordance with gradient descent. During this process, we apply L1 regularization to prevent overfitting by penalizing large weighted coefficients. 
\\ \\
Previous experimental results showed lower levels of correlation between the synthesized and original 12 leads using direct regression over the entire ECG lead. Thus, each lead was individually decomposed into frequency components using Daubechies db4 wavelet decomposition. For each frequency component of the lead being synthesized, a separate regression model is generated. After the regression parameters for each component have been determined, the lead is then reconstructed. Finally, inverse wavelet transform is applied to recombine the reconstructed components to complete a synthesized lead. This process is then repeated for each unknown lead. For example for lead III training:

\subsubsection*{Step 1: Decompose Lead III into Components}
Wavelet decomposition is applied to Lead III, breaking it into multiple components:
\[
\text{III}(t) \xrightarrow{\text{Wavelet Decomposition}} \{A_{\text{III}, L}, D_{\text{III}, 1}, D_{\text{III}, 2}, \dots, D_{\text{III}, L}\}
\]
where:
\begin{itemize}
    \item \(A_{\text{III}, L}\): Approximation coefficients for the low-frequency trends at the highest decomposition level \(L\),
    \item \(D_{\text{III}, l}\): Detail coefficients for high-frequency details at decomposition levels \(l = 1, 2, \dots, L\).
\end{itemize}

\subsubsection*{Step 2: Decompose Known Leads into Components}
The same wavelet decomposition process is applied to the known leads (\(aVR\), \(V2\), \(V5\), \(II\)), producing their respective components:
\[
\{X_{\text{known}, i}\} \xrightarrow{\text{Wavelet Decomposition}} \{A_{\text{known}, L}, D_{\text{known}, 1}, \dots, D_{\text{known}, L}\}
\]
where \(X_{\text{known}, i}\) represents the known leads.

\subsubsection*{Step 3: Differential Reconstruction via Regression}
For each component of Lead III, a regression model is trained using the corresponding components of the known leads to map the known leads to the unknown lead:
\[
\hat{A}_{\text{III}, L} = f_\theta(A_{\text{known}, L})
\]
\[
\hat{D}_{\text{III}, l} = f_\theta(D_{\text{known}, l})
\]
where:
\begin{itemize}
    \item \(f_\theta\): The cubic regression model trained for each component, parameterized by \(\theta\),
    \item \(\hat{A}_{\text{III}, L}\): Reconstructed approximation coefficients for Lead III,
    \item \(\hat{D}_{\text{III}, l}\): Reconstructed detail coefficients for Lead III.
\end{itemize}

\subsubsection*{Step 4: Reconstruct Lead III from Components}
After all components of Lead III are reconstructed, the signal is reassembled using the inverse wavelet transform function:
\[
\hat{\text{III}}(t) = \text{Inverse Wavelet Transform}\left(\hat{A}_{\text{III}, L}, \hat{D}_{\text{III}, 1}, \hat{D}_{\text{III}, 2}, \dots, \hat{D}_{\text{III}, L}\right)
\]

\subsubsection*{Step 5: Repeat for All Missing Leads}
The same process can be systematically applied to reconstruct all other missing leads (\(I, aVL, aVF, V1, V3, V4, V6\)).
\\
This model is called decomposition-based regression. It is preferable to sequence-length reconstruction for physiological signal data that involves critical features that must be preserved. By breaking the signal into frequency components, the model is able to capture sharp changes in the data typically missed by sequence-length regression. In addition, wavelet-based feature extraction is able to preserve certain signal characteristics, such as the QRS complex, a practice that is critical to ECG interpretation. These advantages make the correlation coefficient of decomposition-based regression significantly higher than sequence-length regression. 

\subsection{Classification Algorithm}
The classification algorithm, TAES (Transformer Auto-Encoder System), is a novel transformer-based algorithm that integrates the temporal extraction power of Transformers with the spatial extraction of convolutional networks. An SVM is then used for the classification of the latent-space representation in the middle of the encoder-decoder complex. The encoder and decoder blocks share the same structure in TAES (See: Figure \ref{fig:modeloverview}). The outer layer consists of a transformer and the inner 5 layers consist of progressively shrinking convolutional networks. The outer transformer is able to extract long-distance temporal dependencies, while the inner convolutional layers are able to capture inter-lead spatial relations. In addition, the inner convolutional layers can perform dimensionality reduction using interdispersed max-pooling layers. The model learns by mapping low dimensional features during the encoding process, then reconstructing with these features during decoding to ensure feature accuracy. In this way, the model learns which low-dimensional features are critical to the ECG. Mean Square Error (MSE) is again selected as the loss function for the model, as the goal of TAES training is to reconstruct the input signal as much as possible by reducing error in reconstruction. 

\subsubsection*{Transformer Block}
The transformer model was developed by \citeauthor{DBLP:journals/corr/VaswaniSPUJGKP17}. It is specifically designed for sequence-to-sequence tasks that use self-attention. Given the input \( X \), the Transformer block:
\begin{enumerate}
    \item Computes \( Q, K, V \) to capture inter-dependencies within the sequence:
    \[
    Q = XW_Q, \quad K = XW_K, \quad V = XW_V
    \]
    \item Computes scaled dot-product attention to determine attention:
    \[
    \text{Attention}(Q, K, V) = \text{softmax}\left(\frac{QK^\top}{\sqrt{d_k}}\right)V
    \]
    \item Applies multi-head attention to capture spatial patterns:
    \[
    \text{MultiHead}(Q, K, V) = \text{Concat}(\text{head}_1, \ldots, \text{head}_h)W_O
    \]
    \item Adds residual connection and normalizes to ensure gradient flow:
    \[
    Y = \text{LayerNorm}(X + \text{MultiHead}(Q, K, V))
    \]
    \item Applies the feed-forward network:
    \[
    \text{FFN}(Y) = \text{ReLU}(YW_1 + b_1)W_2 + b_2
    \]
    \item Adds residual connection and normalizes to maintain the gradient:
    \[
    Z = \text{LayerNorm}(Y + \text{FFN}(Y))
    \]
\end{enumerate}

Following the training of the transformer blocks to obtain an initial feature matrix, each layer of the inner convolutional block is trained one by one to reconstruct the mapped features as described by \citeauthor{hettinger_2017_forward}. This model of training ensures the stability of the model and affords us a greater ability to control model parameters. The general convolution operation used by the inside layers is described below.

\subsubsection*{1. Convolution Operation}
For an input feature map \( X \) with dimensions \( H \times W \times C \) (height, width, and channels), a convolution operation applies \( F \) filters \( W_f \) of size \( k \times k \times C \):
\begin{equation}
Y_{i,j,f} = \sum_{m=1}^{k} \sum_{n=1}^{k} \sum_{c=1}^{C} X_{i+m, j+n, c} \cdot W_{f,m,n,c} + b_f
\end{equation}
where:
\begin{itemize}
    \item \( Y_{i,j,f} \): Output value at position \( (i,j) \) for filter \( f \).
    \item \( W_{f,m,n,c} \): Weight of the \( (m,n,c) \)-th element in filter \( f \).
    \item \( b_f \): Bias term for filter \( f \).
    \item \( k \): Kernel size (\( 3 \times 3 \)).
    \item \( C \): Number of input channels.
\end{itemize}

The output feature map \( Y \) has dimensions \( H' \times W' \times F \), where \( H' \) and \( W' \) depend on the stride \( s \) and padding \( p \):
\begin{equation}
H' = \frac{H + 2p - k}{s} + 1, \quad W' = \frac{W + 2p - k}{s} + 1
\end{equation}

\subsubsection*{2. Activation Function (ReLU)}
The activation function introduces non-linearity:
\begin{equation}
\text{ReLU}(z) = \max(0, z)
\end{equation}
where \( z \) is the output of the convolution operation.

The activated feature map \( Y_{\text{ReLU}} \) is:
\begin{equation}
Y_{\text{ReLU}} = \text{ReLU}(Y)
\end{equation}

\subsubsection*{3. Max Pooling}
Max pooling reduces the spatial dimensions of the feature map by applying a pooling window \( p \times p \) (e.g., \( 2 \times 2 \)):
\begin{equation}
Y_{\text{pool},i,j,f} = \max_{m=1}^{p} \max_{n=1}^{p} Y_{\text{ReLU},i+m-1,j+n-1,f}
\end{equation}
where:
\begin{itemize}
    \item \( Y_{\text{pool},i,j,f} \): Output value after pooling at position \( (i,j) \) for filter \( f \).
    \item \( p \): Pooling window size.
\end{itemize}

The resulting dimensions are reduced to:
\begin{equation}
H'' = \frac{H'}{p}, \quad W'' = \frac{W'}{p}
\end{equation}

\subsubsection*{4. Progressive Layer-Wise Transformation}
For each of the 5 convolutional layers, the transformations can be summarized as:
\begin{enumerate}
    \item Convolution:
    \[
    Y_{\text{conv},\ell} = \text{Conv}(X_{\ell-1})
    \]
    \item Activation:
    \[
    Y_{\text{ReLU},\ell} = \text{ReLU}(Y_{\text{conv},\ell})
    \]
    \item Pooling:
    \[
    X_{\ell} = \text{MaxPool}(Y_{\text{ReLU},\ell})
    \]
\end{enumerate}
where \( \ell \) is the layer index, and \( X_0 \) is the input to the first layer.

\subsubsection*{Example of Dimensionality Reduction}
Assume the input is a 12-lead ECG beat with 100 data points per lead:
\[
X \in \mathbb{R}^{100 \times 12}
\]
where:
\begin{itemize}
    \item \( 100 \): Sequence length (time points).
    \item \( 12 \): Number of leads.
\end{itemize}

Each layer progressively reduces the dimensionality as follows:

\subsubsubsection*{Layer 1}
- **Input size**: \( 100 \times 12 \)
- **Convolution**:
  - Filters: \( 32 \)
  - Kernel size: \( 3 \)
  - Stride: \( 1 \)
  - Padding: Same
- **Output size after convolution**: \( 100 \times 32 \)
- **Max pooling**:
  - Pooling size: \( 2 \)
  - Stride: \( 2 \)
- **Output size after pooling**:
\[
50 \times 32
\]

\subsubsubsection*{Layer 2}
- **Input size**: \( 50 \times 32 \)
- **Convolution**:
  - Filters: \( 64 \)
  - Kernel size: \( 3 \)
  - Stride: \( 1 \)
  - Padding: Same
- **Output size after convolution**: \( 50 \times 64 \)
- **Max pooling**:
  - Pooling size: \( 2 \)
  - Stride: \( 2 \)
- **Output size after pooling**:
\[
25 \times 64
\]

\subsubsubsection*{Layer 3}
- **Input size**: \( 25 \times 64 \)
- **Convolution**:
  - Filters: \( 128 \)
  - Kernel size: \( 3 \)
  - Stride: \( 1 \)
  - Padding: Same
- **Output size after convolution**: \( 25 \times 128 \)
- **Max pooling**:
  - Pooling size: \( 2 \)
  - Stride: \( 2 \)
- **Output size after pooling**:
\[
12 \times 128
\]

\subsubsubsection*{Layer 4}
- **Input size**: \( 12 \times 128 \)
- **Convolution**:
  - Filters: \( 256 \)
  - Kernel size: \( 3 \)
  - Stride: \( 1 \)
  - Padding: Same
- **Output size after convolution**: \( 12 \times 256 \)
- **Max pooling**:
  - Pooling size: \( 2 \)
  - Stride: \( 2 \)
- **Output size after pooling**:
\[
6 \times 256
\]

\subsubsubsection*{Layer 5}
- **Input size**: \( 6 \times 256 \)
- **Convolution**:
  - Filters: \( 512 \)
  - Kernel size: \( 3 \)
  - Stride: \( 1 \)
  - Padding: Same
- **Output size after convolution**: \( 6 \times 512 \)
- **Max pooling**:
  - Pooling size: \( 2 \)
  - Stride: \( 2 \)
- **Output size after pooling**:
\[
3 \times 512
\]
where:
\begin{itemize}
    \item \( 3 \): Reduced sequence length.
    \item \( 512 \): Number of feature channels.
\end{itemize}

After all layers are trained, the model is then reconstructed layer by layer. The final model produces a low-dimensional feature map as the latent space representation. This low-dimensional representation of the original ECG beat is then fed through a One vs. One (OvO) Support Vector Machine (SVM). The classic SVM classifies data by maximizing the separation of two sets of data in the hyperplane. In this case, non-linearity must be introduced as the data is rather complex and is not separable via linear lines, thus the RBF kernel is utilized. In essence, the data is mapped into infinite-dimensional feature space to introduce possible planes of separation, aka the kernel trick. Furthermore, since there are multiple classes (6), 15 separate binary classifiers are established to classify the test samples, each classifier is then compared and aggregated to determine the model classification. Grid search is used to determine the optimal value for the coefficient C. 

\subsubsection*{RBF Kernel}
The RBF kernel between two data points \( \mathbf{x}_i \) and \( \mathbf{x}_j \) is defined as:
\[
K(\mathbf{x}_i, \mathbf{x}_j) = \exp\left(-\gamma \|\mathbf{x}_i - \mathbf{x}_j\|^2\right)
\]
where:
\begin{itemize}
    \item \( \mathbf{x}_i, \mathbf{x}_j \in \mathbb{R}^d \): Data points in the input space with dimensionality \( d \).
    \item \( \|\mathbf{x}_i - \mathbf{x}_j\|^2 \): Squared Euclidean distance:
    \[
    \|\mathbf{x}_i - \mathbf{x}_j\|^2 = \sum_{k=1}^{d} (x_{i,k} - x_{j,k})^2
    \]
    \item \( \gamma > 0 \): Kernel parameter controlling the influence of individual training samples.
\end{itemize}

\subsubsection*{Feature Space Mapping}
The RBF kernel implicitly maps the input data into an infinite-dimensional feature space \( \phi(\mathbf{x}) \), such that:
\[
K(\mathbf{x}_i, \mathbf{x}_j) = \langle \phi(\mathbf{x}_i), \phi(\mathbf{x}_j) \rangle
\]
where \( \langle \cdot, \cdot \rangle \) denotes the inner product in the feature space.

\subsubsection*{SVM Decision using RBF}
The SVM decision function using the RBF kernel is given by:
\[
f(\mathbf{x}) = \sum_{i=1}^n \alpha_i y_i K(\mathbf{x}_i, \mathbf{x}) + b
\]
where:
\begin{itemize}
    \item \( \alpha_i \): Lagrange multipliers.
    \item \( y_i \in \{-1, 1\} \): Class labels of training data.
    \item \( K(\mathbf{x}_i, \mathbf{x}) \): RBF kernel.
    \item \( b \): Bias term.
\end{itemize}

\subsubsection*{One vs. One SVM}
The One-vs-One (OvO) strategy extends the binary SVM model to multi-class classification. For a dataset with \( K \) classes, \( C_1, C_2, \dots, C_K \), the OvO approach constructs:
\[
\binom{K}{2} = \frac{K(K-1)}{2}
\]
binary classifiers, where each classifier distinguishes between two classes \( C_i \) and \( C_j \) (\( i \neq j \)). In this case, 15 classifiers are constructed, as 6 classes are being classified. 

\textbf{Construction of Binary Classifiers} \\
For each pair of classes \( (C_i, C_j) \):
\begin{itemize}
    \item Extract all training samples belonging to \( C_i \) and \( C_j \).
    \item Assign binary labels:
    \[
    y = 
    \begin{cases} 
    +1, & \text{if the sample belongs to } C_i, \\
    -1, & \text{if the sample belongs to } C_j.
    \end{cases}
    \]
    \item Train an SVM classifier \( h_{ij}(\mathbf{x}) \) to distinguish between \( C_i \) and \( C_j \).
\end{itemize}

The decision function for the binary classifier is then denoted as:
\[
h_{ij}(\mathbf{x}) = \text{sign}\left(f_{ij}(\mathbf{x})\right),
\]
where:
\[
f_{ij}(\mathbf{x}) = \sum_{k=1}^n \alpha_k y_k K(\mathbf{x}_k, \mathbf{x}) + b_{ij},
\]
and:
\begin{itemize}
    \item \( \alpha_k \): Lagrange multipliers for the support vectors.
    \item \( y_k \): Binary label of the \( k \)-th training sample.
    \item \( K(\mathbf{x}_k, \mathbf{x}) \): Kernel function (e.g., RBF kernel).
    \item \( b_{ij} \): Bias term.
\end{itemize}

\textbf{Classification of Test Samples} \\
For a test sample \( \mathbf{x} \), the class with the highest number of votes across all pairwise classifiers is assigned to the test sample:
\[
\hat{y} = \arg\max_{i} \sum_{j \neq i} \mathbf{1}\left(h_{ij}(\mathbf{x}) = +1\right),
\]
where:
\begin{itemize}
    \item \( \mathbf{1}(\cdot) \): Indicator function that equals 1 if the condition is true, and 0 otherwise.
    \item \( h_{ij}(\mathbf{x}) = +1 \): Indicates a vote for class \( C_i \) from the \( (i, j) \)-th classifier.
\end{itemize}

A summary of the overall classification model, TAES, is presented below in Figure \ref{fig:modeloverview}. 

\begin{figure} [!h]
    \centering
    \includegraphics[width=\linewidth]{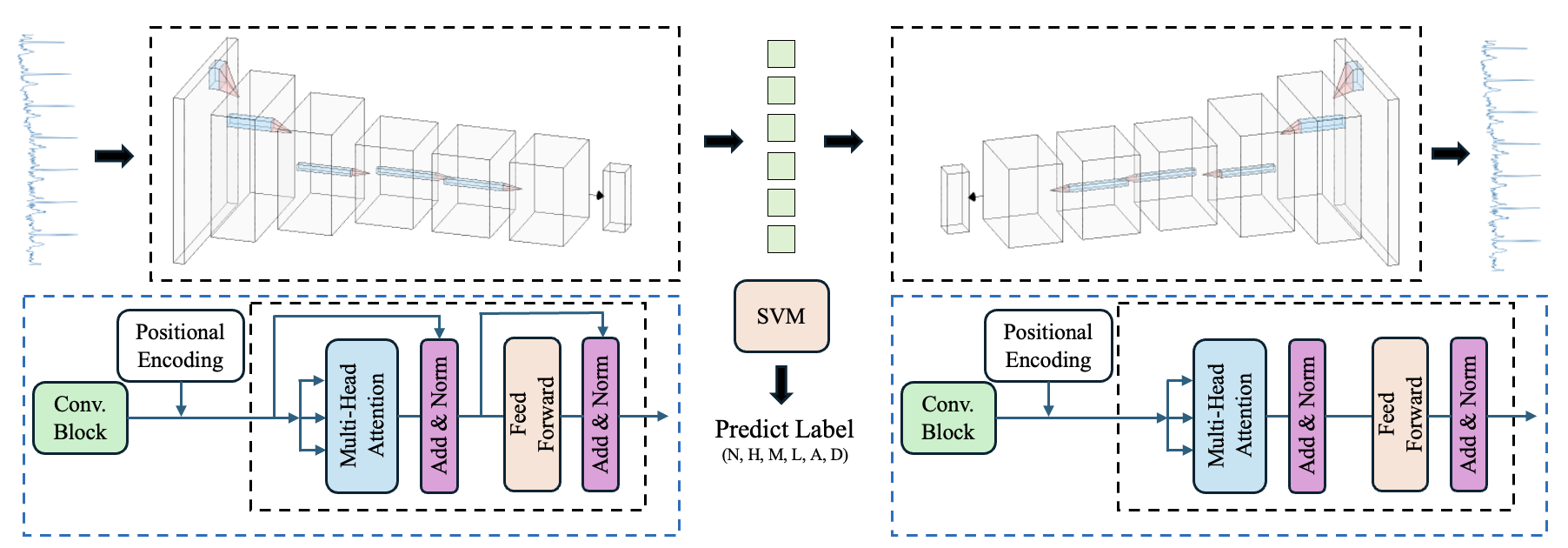}
    \caption{Overview of Transformer Auto-Encoder System. In the model, the transformer serves as the outer layer of the model. Internally, 5 convolutional autoencoders extract spatial information. The model is trained by reconstructing ECGs to learn the features mapped by the model. In this way, the model learns which features are important by reconstructing the ECG and determining which type of feature has the greatest necessity in reconstruction. The latent state representation, following a pass through the encoder block, is taken and fed into the OvO SVM model before a final classification is determined by the model.}
    \label{fig:modeloverview}
\end{figure}

\newpage
\subsection{Human Subject Trial Design}
Our population cohort consisted of 30 individuals in the Pittsburgh, Pennsylvania area. The IRB/Ethics Committee of Shady Side Academy gave ethical approval for this work. The IRB consisted of an administrator, teacher, RN, MD, psychologist, and professional with a research focus in cardiology. No members of the IRB were personally related to either author. All participants consented via a consent form developed by the authors in conjunction with the IRB. The anonymized data that supports the findings of this study is available from the corresponding author, E.X., upon reasonable request. Aggregate participant data is presented here. All data was utilized for validation of the smartwatch protocol. A subgroup of 20 subjects was selected for diagnostic validation. 10 subjects had known cardiac conditions prior to the study and were pre-selected for the diagnostic validation group. We recognize this as a fundamental limitation of our study. Following this, 10 additional subjects were randomly selected for the control group. 
\\
This random group (n = 20) of patients was screened by a licensed physician using the smartwatch protocol and then processed through our CSS for diagnostic labeling. We compared the physician label and algorithmic label to determine agreement. Aggregate data is displayed as a confusion matrix. 

\subsubsection{Inclusion and Exclusion Criteria}
The following criteria were offered for inclusion in our study:
\begin{enumerate}
    \item Participants must be above the age of 13 and below the age of 60 as of 02/01/2024.
    \item Participants must be willing and able to follow the given protocol. 
    \item Participants must be able to be administered a 12-lead ECG. They must be able to remain still for the recording period. 
\end{enumerate}

The following criteria were presented for exclusion from our study:
\begin{enumerate}
    \item Systolic BP <100 mmHg 
    \item Diastolic BP <60 mmHg
    \item Tremors, Situs Invertus, or any other condition that may cause incorrect lead placement or that may increase the noise of the ECG trace.   
\end{enumerate}

\subsubsection{Device Parameters}
The GE MAC1200 Electrocardiogram machine was used to measure the 12-lead ECG. The paper was fed at a rate of 25 mm/s at a 1,000 Hz sampling rate. All ECG signals were recorded by a medical professional in the supine position for 2 minutes. The smartwatch protocol was performed via an Apple Watch S7 which records a 1 lead ECG at 25 mm/s, 250 Hz, and 10mm/mv. All leads were recorded via the protocol described in Figure \ref{protocol} in the supine position. The 4 lead ECG was recorded immediately following the 12 lead ECG. Each lead was recorded sequentially for 30 seconds. All participants were provided the same instruction sheet for performing the ECG with a medical professional assisting if needed. All ECGs were exported from the Apple Health app via the PDF export function and converted into raw data via the methodology and software prescribed by \citeauthor{fortune_2022_digitizing}.  

\section{Statistical Analysis}

Statistical analysis was performed using IBM SPSS Statistics v28 and R. Quantitative data is displayed in the format mean $\pm$ standard deviation. Observational data is presented textually. In all tests, a p-value < 0.05 is considered statistically significant. All other data is compared directly to calculate the correlation coefficient. All data with a correlation coefficient greater than 70\% is assumed to be significantly correlated between both ECGs. Bland Altman analysis is used to quantify and graphically show agreement between the real 12 lead ECG and the smartwatch ECG for each of the described criteria below:

\begin{enumerate}
    \item QRS Complex: Amplitude(mV), Duration(ms), and Morphology(+/-)
    \item P Wave: Amplitude(mV), Duration(ms), and Morphology(+/-)
    \item PR Interval: Duration(ms)
    \item T Wave: Amplitude(mV) and Morphology(+/-)
    \item QTc (QT Corrected) Interval: Duration(ms) via Bazett's Formula
    \item ST Morphology: Isoelectric, Elevated, or Depressed 
\end{enumerate}

\section{Results}
\subsection{Regression-Based Upscaling}
\begin{figure} [!h]
    \centering
    \includegraphics[width=0.65\linewidth]{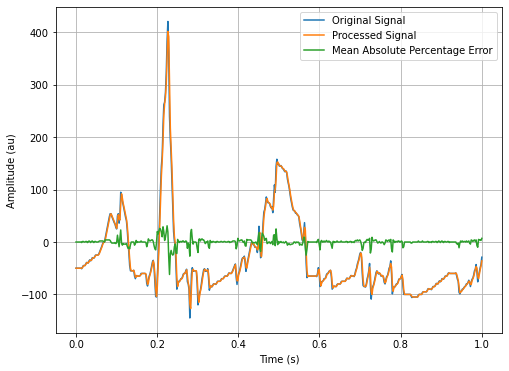}
    \caption{Example of ECG Synthesis where the blue line denotes the original signal, the orange line denotes the processed signal, and the green line depicts the MAPE.}
    \label{fig:synth}
\end{figure}

\begin{figure} [!h]
    \centering
    \includegraphics[width=0.8\linewidth]{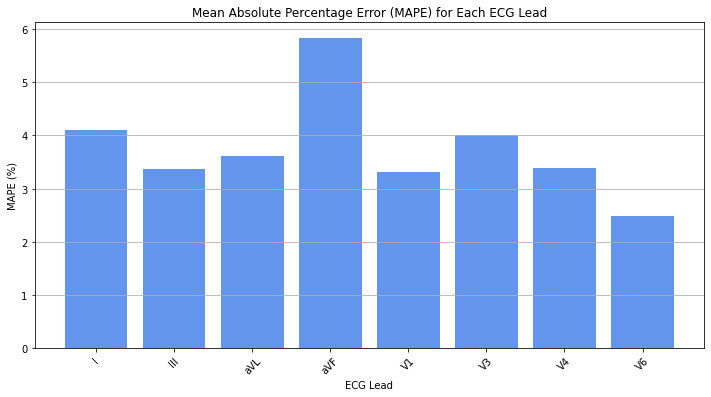}
    \caption{Aggregate MAPE Values evaluated across the entirety of the training database.}
    \label{fig:mape}
\end{figure}

Figure \ref{fig:synth} shows an example of a synthesized ECG beat using the decomposition-based regression methodology. In order to evaluate the difference between the synthesized beat and the actual beat, the Mean Absolute Percent Error (MAPE) is calculated according to the formula: 
\[
\text{MAPE} = \frac{1}{n} \sum_{i=1}^n \left| \frac{y_i - \hat{y}_i}{y_i} \right| \times 100
\]
An aggregate chart showing MAPE values for all synthesized leads across the entire database is shown in Figure \ref{fig:mape}. The db4 wavelet decomposition method was used to decompose the lead into smaller elements. 

\subsection{Classification Algorithm}
\begin{figure} [!h]
    \centering
    \includegraphics[width=0.7\linewidth]{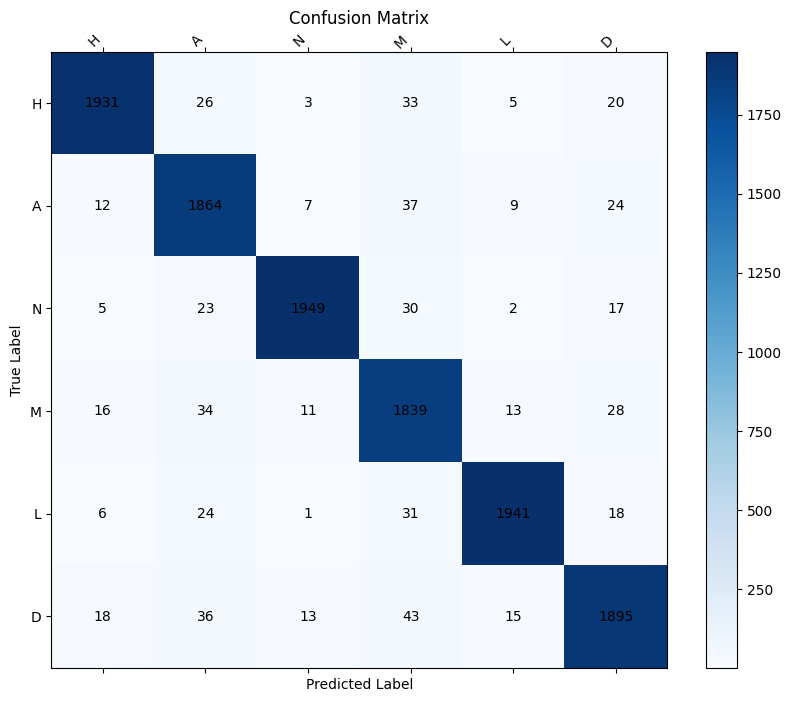}
    \caption{Confusion Matrix for Labeling. H = hypertrophic cardiomyopathy, A = arrhythmogenic left ventricular hypertrophy, N = normal sinus rhythm, M = myocarditis, L = long QT syndrome, D = dilated cardiomyopathy. Darker blue squares indicate higher number of individuals within that category.}
    \label{fig:classificationoverall}
\end{figure}

A confusion matrix was generated for our classification model on the upscaled testing set, shown in Figure \ref{fig:classificationoverall}. The overall evaluated model statistics are available in Table \ref{tab:classmodel}. Macro and micro averages in Table \ref{tab:macmic}. High accuracy is specifically shown in the classification of hypertrophic cardiomyopathy (HCM). Our deep layers effectively allow us to extract features that are typically not seen on HCM ECGs by physicians, reducing the 15 to 20\% of cases in which HCM cannot be directly observed. Results here demonstrate that beat-based feature extraction classification through TAES has high accuracy versus physician interpretation and contemporary ML models. In addition, we present sensitivity and specificity results in Figure \ref{tab:CG_comp}.

\begin{table*}[!h]
	\caption{Classification Model Statistics per Disorder - Test Set}
        \label{tab:classmodel}
	\centering 
        \small
	\begin{tabular}{@{}ccccc@{}}
		\toprule
		\cmidrule(r){2-4}
		Disorder & Precision & Recall & F1 Score  \\
		\midrule
		Hypertrophic Cardiomyopathy (H) & 97.2\% & 95.6\% & 0.96 \\
		Arrhythmogenic Left Ventricular Hypertrophy (A) & 92.9\% & 95.41\% & 0.94 \\
		Normal Sinus Rhythm (N) & 98.2\% & 96.1\% & 0.97 \\
            Myocarditis (M) & 91.3\% & 94.7\% & 0.93\\
            Long QT Syndrome (L) & 97.8\% & 96.1\% & 0.97\\
            Dilated Cardiomyopathy (D) & 94.7\% & 93.4\% & 0.94\\
		\bottomrule
	\end{tabular}
\end{table*}

\begin{table*}[!h]
	\caption{Macro and Micro Averages on Classification Algorithm}
        \label{tab:macmic}
	\centering 
        \small
	\begin{tabular}{p{0.5\linewidth} p{0.1\linewidth} p{0.1\linewidth}}
		\toprule
		Model & Macro F1 & Micro F1 \\
		\midrule
		Transformer Auto-Encoder System (TAES) & 0.95 & 0.97\\
		\bottomrule
	\end{tabular}
\end{table*}

The performance metrics are defined as follows:
\[
\text{Precision} = \frac{\text{TP}}{\text{TP} + \text{FP}}, \quad
\text{Recall (Sensitivity)} = \frac{\text{TP}}{\text{TP} + \text{FN}}, \quad
\text{Specificity} = \frac{\text{TN}}{\text{TN} + \text{FP}}, \quad
\text{F1 Score} = 2 \cdot \frac{\text{P} \cdot \text{R}}{\text{P} + \text{R}}
\]
where:
\begin{itemize}
    \item P = Precision
    \item R = Recall
\end{itemize}

\subsubsection{Model Parameters}
The model was tested with various hyperparameter combinations. It is at optimal when the training epochs of each layer is 24, the batch size is 32, and the optimizer is Adam with a learning rate of 0.01. The transformer block’s parameters are defined as dmodel (64), head (3), and layer (4). A radial base function kernel is used for SVM with a penalty coefficient C of 0.5. The overall dimensions of the model are summarized in Table \ref{tab:modelstruc}. In addition, for a 1-second long ECG signal on the completed TAES model, preprocessing takes 1.213 ms on the CPU, neural network inference takes 0.052 ms on the GPU, and SVM prediction takes 0.441 ms on the CPU. The time from signal acquisition to neural network prediction is 1.706 ms, demonstrating that TAES may be used for high throughput diagnostics. 
\begin{table}[h!]
\centering
\caption{Shape of Data through the TAES Model}
\begin{tabular}{@{}clccc@{}}
\toprule
\textbf{Layer} & \textbf{Type}        & \textbf{Input size}        & \textbf{Output Size} & \textbf{Feature Num} \\ \midrule
1                 & Transformer         & batch size*360*1           & batch size*360                   & 360                  \\
2                 & Conv1         & batch size*360             & batch size*252                   & 252                  \\
3                 & Conv2         & batch size*252             & batch size*128                   & 128                  \\
4                 & Conv3         & batch size*128             & batch size*72                    & 72                   \\
5                 & Conv4         & batch size*72              & batch size*56                    & 56                   \\ 
6                 & Conv5         & batch size*56              & batch size*32                    & 32                   \\ \bottomrule

\end{tabular}
\label{tab:modelstruc}
\end{table}

\subsection{Human Subject Testing}
Table \ref{tab:pop} shows the population demographics of our testing cohort. Duration, amplitude, and morphology statistics are reported in Table \ref{tab:ecg_comparison} and \ref{tab:ecg_comparison2} along with correlation coefficient calculations to determine the agreement between the real ECG and watch protocol ECG. There was no significant difference between the commercial 12-lead ECG and the sequential Apple Watch S7 method (p < 0.05) in terms of feature replication. Bland Altman Analysis was further used in Supplementary Figures S1-12 to clarify and confirm differences/agreement in the two methods of measurement. These showed no statistically significant difference between both methodologies. \\ \\
Morphology results are reported here as only three signals included a measurement. The +/- morphology of the T Wave and P Wave differed in none of the screened individuals between the standard ECG and the Apple Watch ECG. The morphology of the ST segment was also consistent between the standard and Apple Watch. This data was not reported in the figures below but was observed in ECG interpretation. \\ \\
The diagnostic subgroup of 20 individuals (10 afflicted, 10 control) was subject to physician interpretation and CSS interpretation. Aggregate data is presented as a confusion matrix below in Figure \ref{fig:aggregatedata}. 

\begin{table*} [!h]
	\caption{Population Statistics}
        \label{tab:pop}
	\centering 
	\begin{tabular}{ c c } 
		\toprule
		  Variables and Quantity \\
		\midrule
		Age & 16.2 $\pm$ 1 years old\\
            Sex (Male/Female) & 16/14 \\
            Systolic Blood Pressure & 121 $\pm$ 13 mmHg \\
            Diastolic Blood Pressure & 73 $\pm$ 11 mmHg \\
            Average Beats Per Minute & 74 $\pm$ 6 \\
		\bottomrule
	\end{tabular}
\end{table*}

\begin{table}[!h]
  \centering
  \caption{Comparison of ECG duration in milliseconds from standard versus smartwatch ECG devices. All data presented here is the aggregate of all 30 patients in the validation group. The correlation coefficient for all duration comparisons is above 0.9, showing that the smartwatch protocol is effective for taking ECG.}
\resizebox{\textwidth}{!}{
  \begin{tabular}{lccccc}
    \toprule
    \textbf{Criterion} & \textbf{Lead} & \textbf{Standard ECG} & \textbf{Smartwatch ECG} & \textbf{Correlation Coefficient} & \textbf{p-Value} \\
    \midrule
    P Wave   & II  & 77.34 $\pm$ 11.75 & 76.58 $\pm$ 11.67 & 0.94 & <0.001 \\
             & AvR & 77.29 $\pm$ 13.12 & 77.88 $\pm$ 13.21 & 0.94 & <0.001 \\
             & V2  & 77.47 $\pm$ 14.63 & 78.69 $\pm$ 13.95 & 0.96 & <0.001 \\
             & V5  & 77.52 $\pm$ 14.70 & 78.72 $\pm$ 13.90 & 0.95 & <0.001 \\
    \midrule
    PR Interval & II  & 172.35 $\pm$ 24.60 & 171.42 $\pm$ 24.68 & 0.98 & <0.001 \\
                & AvR & 173.78 $\pm$ 30.01 & 172.75 $\pm$ 29.20 & 0.97 & <0.001 \\
                & V2  & 172.15 $\pm$ 28.30 & 172.69 $\pm$ 26.55 & 0.98 & <0.001 \\
                & V5  & 172.20 $\pm$ 28.25 & 172.65 $\pm$ 26.50 & 0.97 & <0.001 \\
    \midrule
    QRS Complex & II  & 87.48 $\pm$ 11.74  & 87.68 $\pm$ 11.20  & 0.94 & <0.001 \\
                & AvR & 88.83 $\pm$ 10.76  & 89.04 $\pm$ 10.49  & 0.93 & <0.001 \\
                & V2  & 87.15 $\pm$ 14.55  & 87.28 $\pm$ 14.01  & 0.95 & <0.001 \\
                & V5  & 87.18 $\pm$ 14.52  & 87.25 $\pm$ 13.98  & 0.94 & <0.001 \\
    \midrule
    QT Interval & II  & 369.67 $\pm$ 44.25 & 369.92 $\pm$ 44.30 & 0.99 & <0.001 \\
                & AvR & 368.48 $\pm$ 42.23 & 367.47 $\pm$ 40.65 & 0.98 & <0.001 \\
                & V2  & 369.57 $\pm$ 42.35 & 369.20 $\pm$ 42.17 & 0.99 & <0.001 \\
                & V5  & 369.60 $\pm$ 42.40 & 369.23 $\pm$ 42.15 & 0.98 & <0.001 \\
    \midrule
    T Wave & II  & 132.44 $\pm$ 24.77 & 133.14 $\pm$ 24.83 & 0.97 & <0.001 \\
            & AvR & 132.81 $\pm$ 24.46 & 133.17 $\pm$ 23.95 & 0.97 & <0.001 \\
            & V2  & 133.79 $\pm$ 24.18 & 133.84 $\pm$ 23.19 & 0.98 & <0.001 \\
            & V5  & 133.74 $\pm$ 24.14 & 133.80 $\pm$ 23.16 & 0.96 & <0.001 \\
    \bottomrule
  \end{tabular}
  }
  \label{tab:ecg_comparison}
\end{table}

\begin{table}[!h]
  \centering
  \caption{Comparison of ECG amplitude in millivolts between standard and smartwatch ECG devices across different leads. All data in this graph is an aggregate of the 30-person validation group. The correlation coefficient for almost all values (except for P wave, Lead II) is above 0.8, showing that there exists a high degree of association between the standard vs. smartwatch devices.}
\resizebox{\textwidth}{!}{
  \begin{tabular}{lccccc}
    \toprule
    \textbf{Criterion} & \textbf{Lead} & \textbf{Standard ECG} & \textbf{Smartwatch ECG} & \textbf{Correlation} & \textbf{p-Value} \\
    \midrule
    P Wave   & II  & 0.22 ± 0.07 & 0.23 ± 0.06 & 0.78 & <0.001 \\
             & AvR & 0.23 ± 0.10 & 0.24 ± 0.09 & 0.86 & <0.001 \\
             & V2  & 0.21 ± 0.08 & 0.22 ± 0.07 & 0.83 & <0.001 \\
             & V5  & 0.20 ± 0.09 & 0.21 ± 0.08 & 0.88 & <0.001 \\
    \midrule
    QRS Complex & II  & 1.88 ± 0.69 & 1.90 ± 0.67 & 0.93 & <0.001 \\
                & AvR & 1.92 ± 0.75 & 1.94 ± 0.73 & 0.91 & <0.001 \\
                & V2  & 1.85 ± 0.68 & 1.87 ± 0.70 & 0.95 & <0.001 \\
                & V5  & 1.90 ± 0.65 & 1.92 ± 0.66 & 0.89 & <0.001 \\
    \midrule
    T Wave & II  & 0.30 ± 0.14 & 0.31 ± 0.15 & 0.89 & <0.001 \\
            & AvR & 0.29 ± 0.16 & 0.30 ± 0.17 & 0.85 & <0.001 \\
            & V2  & 0.28 ± 0.15 & 0.29 ± 0.16 & 0.90 & <0.001 \\
            & V5  & 0.27 ± 0.17 & 0.28 ± 0.18 & 0.87 & <0.001 \\
    \bottomrule
  \end{tabular}
  }
  \label{tab:ecg_comparison2}
\end{table}

\begin{figure} 
    \centering
    \includegraphics[width=0.75\linewidth]{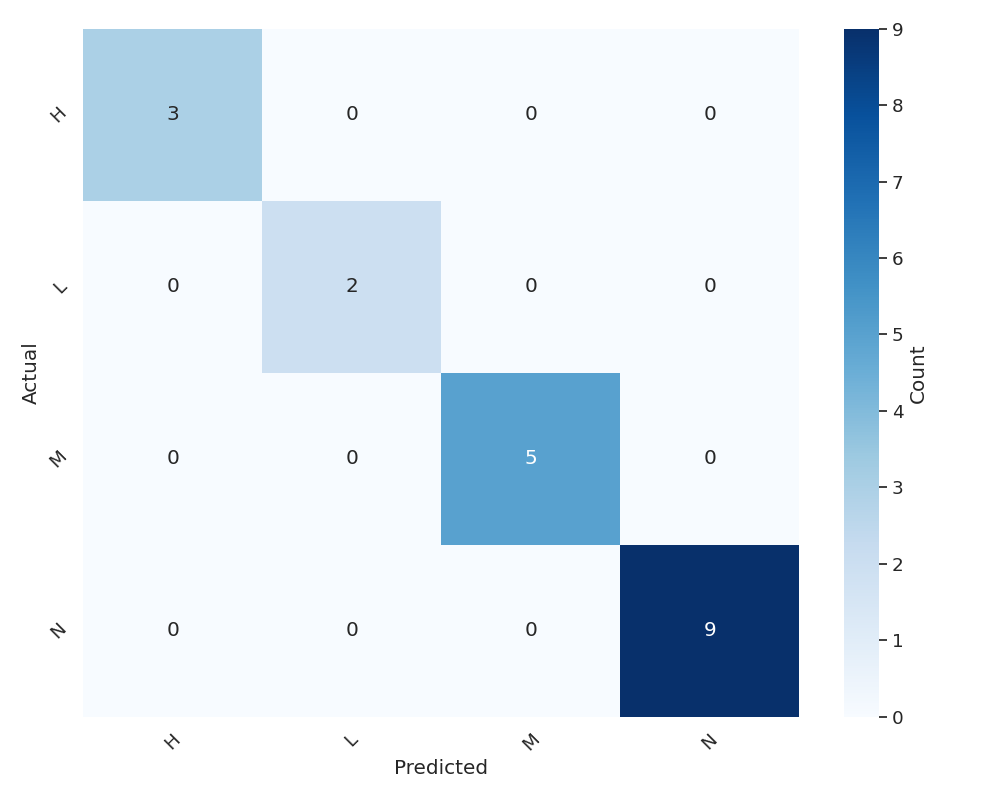}
    \caption{Aggregate Confusion Matrix of Human Subject Data. Darker colors indicate a greater number of individuals.}
    \label{fig:aggregatedata}
\end{figure}

\newpage
\section{Discussion}
The results of this study support the feasibility of the proposed comprehensive screening system (CSS) for high throughput detection of cardiac risk factors. Our findings demonstrate the accuracy of our protocol compared to current solutions, potentially decreasing the burden of ECG screening on healthcare infrastructure and drawing pre-athletic ECG screening into the realm of possibility. 
\begin{table} [!h]
	\caption{Proposed CSS Compared with Existing Pre-Participation Evaluation (PPE)}
	\label{tab:CG_comp1}
	\centering
	\begin{tabular}{lccc}
		\toprule
		\textbf{Criterion} & \textbf{14-point PPE} & \textbf{12-lead ECG} & \textbf{This work} \\
		\midrule
		Sensitivity & 18.8\% & 94\% & 97.3\% \\
        Specificity  & 68.0\% & 93\% & 99.1\%\\
        FPR & 32.0\% & 7\% & 2\% \\
        Price & N/A & \$>1,000 \cite{USA}  & \$399 \\
        Require physician? & Yes & Yes & No \\
		\bottomrule
	\end{tabular}  
\end{table}

\subsection{Screening Methodology}

Our SW screening method showed minimal difference compared to the gold standard commercial 12-lead ECG in terms of the correlation coefficient. All duration measurements had a CC value of >0.9 and nearly all amplitude measurements had a value of >0.8, demonstrating that the protocol used for watch screening is accurate. The high correlation of our SW ECG method surpassed \citeauthor{Touiti2023} and \citeauthor{Behzadi2020}. Bland Altman Analysis of both amplitude and duration (Supplementary Figures S1-12) showed high agreement between both screening methodologies. The bias line for QRS, QT, T, and PR duration was within 0.3 of 0 bias, indicating high agreement in these segments of the ECG. P wave bias was also generally low, with a mild outlier in Lead II. The distribution was mostly random with no observable trend indicating that the difference between the two methods did not change with the duration (ms) of the observed segment. Additionally, all points were within the limits of agreement (bias $\pm$ 1.96 $\times$ SD) showing high agreement between both methods. A large spectrum of points was also present, consistent with a large spectrum of physiological signals being recorded. We noticed similar trends with our amplitude plots, where most values were within the limits of agreement. These results confirm the general viability of our SW protocol, as our current testing indicated that the watch generally agrees with the 12 lead gold standards on critical diagnostic metrics. This methodology may be applied to high-throughput screening scenarios, where the watch provides a cost-effective device solution. Table \ref{tab:CG_comp1} shows that the cost of a 12 lead ECG is greater than 1000 USD, while the cost of our screening system is only the 399 USD cost of an Apple Watch S7. Using the Apple Watch significantly reduces cost and the complexities associated with lead placement that significantly hinder the full 12 lead ECG \cite{Rjoob2020}. The upscaling algorithm also shows mean absolute percent error values of <10\% on all synthesized leads. Most leads are below 5\%. This error in synthesis is similar to that achieved by \citeauthor{Sohn2020}. Thus, our upscaling algorithm can synthesize complete ECGs that nearly replicate clinical gold-standard ECGs. 

\subsection{Combined Algorithm}

In the beat-based classification algorithm, the sensitivity, specificity, and F1 score of our methodology are 95.33\%, 99.12\%, and 97.0\%, respectively. These values demonstrate that the TAES algorithm is viable for high-throughput screening, as they significantly surpass the benchmark for physician-interpreted ECGs. In addition, we show that the time to interpretation on the trained TAES model is <2ms, further demonstrating that our algorithm is feasible for high throughput screening. Table \ref{tab:CG_comp1} reveals that our method exceeded the 14-point PPE and the traditional 12-lead ECG in terms of sensitivity, specificity and cost. Our findings show that the typical physician-interpreted ECG is not necessary for accurate results, showing that the usage of ML in classification could surpass the abilities of a human physician. TAES can facilitate easy-to-use automated screening. Similar recent studies are shown in comparison with our work in Table \ref{tab:CG_comp}. Our sensitivity and specificity values were consistent alongside SOTA (State-of-the-Art) models for 12-lead ECG. Our model is able to classify upscaled ECG traces with nearly the same accuracy as SOTA algorithms. The TAES algorithm extracts spatial and temporal features within the dataset, making it particularly useful for the diagnosis of structural disorders (e.x cardiomyopathies) where certain areas of the heart become inflamed. These disorders are difficult to detect through physician interpretation, thus using TAES could mitigate this concern. In addition, utilizing TAES within our CSS allows us to mitigate any issues regarding physician fatigue and consistency in interpretation. The TAES algorithm can achieve similar results to the current SOTA while using upscaled ECG signals. In the future, we will experiment with full 12 lead signals to see if this lends itself to any tangible advantage. As a system, the upscaling algorithm, watch, and TAES protocol show great promise as a potential CSS for mass screening. Not only is the watch method significantly more cost-effective, it also has similar accuracy to real ECGs and results in clear ECG signals for interpretation. Furthermore, TAES extracts deeper features of the ECG not typically observed by human physicians, allowing it to diagnose structural disorders with greater efficacy than nearly all other algorithms and methods of interpretation. Our CSS has the potential to be utilized for mass screening in the future. 

\begin{table*} 
  \caption{Comparison with the results of similar studies}
  \centering
  \begin{tabular}{lccccc}
    \toprule
    \textbf{Method} & \textbf{Model} & \textbf{Se (\%)} & \textbf{Sp (\%)} & \textbf{F1 (\%)}\\ 
    \midrule
    \citeauthor{Smigiel2021} & CNN  & ~80 & ~80 & / \\

    \citeauthor{Ding2023} & Transformer Auto-Encoder  & 98.84 & 99.84 & 99.13 \\

    \citeauthor{Oh2019} & U-Net \& CNN & 94.44 & 98.26 & / \\

    \citeauthor{Thomas2015} & Complex Wavelet Transform \& ANN & 94.44 & 98.26 & / \\

    \citeauthor{Oh2018} & LSTM \& CNN & 97.50 & 98.70 & / \\

    \citeauthor{Meng2022} & Transformer with LightConv Attention & 92.44 & / & 93.63 \\

    \citeauthor{Hu2022} & Transformer-Based NN & 92.51 & 99.84 & 93.88 \\

    \citeauthor{Han2022} & Convolutional Transformer & 98.10 & / & / \\

    \citeauthor{Hou2020} & LSTM-based Auto-Encoder & 99.35 & 99.84 & / \\
    
    \textbf{This work}  & Transformer Auto-Encoder  & 95.33 & 99.12  & 97.00 \\

    \bottomrule
  \end{tabular}
  \label{tab:CG_comp}
  
\end{table*}

\subsection{Limitations}

In this study, the algorithms used for upscaling and classification were trained on datasets that only consisted of the most common 5 contributory disorders to Sudden Cardiac Arrest (SCA). Our preliminary comprehensive screening system (CSS) is not able to detect all contributory disorders to SCA from ECG signals, thus, future research can focus on incorporating more diverse and representative datasets. Although most participants were identified with the correct label, further testing on a larger population with more contributory disorders is necessary to prove our methodology's clinical accuracy in the detection of diseased labels (H, A, M, L, D) is not accidental. In addition, our algorithms are "black boxes" and as such we do not know if they hold any implicit biases or if they can diagnose disorders across all people. The majority of our disease group (n=10) was male (n=8), thus the algorithm might struggle with female patients and we would not know. Furthermore, the dataset was not racially stratified, nor was our testing group. It would be critical to test the efficacy of our algorithm in a more diverse setting, to ensure the general usefulness of our system for all. Finally, in the detection of other disorders, specifically more structural disorders, there may be issues with the decomposition-based regression style of upscaling. In this system, each signal is essentially picked apart and the individual frequency components are used for regression. However, this system may face issues regarding the fine preservation of details. While it is preferable to sequence-length regression, there still may be degradation of the important components of the signal that could lead to misclassifications. These issues will be addressed in future studies.

\section{Conclusion}

This paper proposes a preliminary comprehensive screening system (CSS) for the early detection of disorders that present as risk factors for SCA in athletes. A protocol was developed for Apple Watch-based 4-lead ECG. Furthermore, a new type of regression, decomposition regression was created and used to synthesize a 12-lead ECG from the limited set of 4 leads. Then, TAES (Transformer Auto-Encoder System) was created and used to implement the classification of the ECGs, making our CSS a comprehensive system. The results show baseline viability in our smartwatch extraction protocol and our diagnostic subgroup shows promising results with no misidentifications within a 20-subject cohort. The developed CSS is straightforward, intuitive, and easy to use, potentially improving patient outcomes. Future work focuses on the improvement of the diagnostic algorithm, as well as larger human subject trials to determine true CSS accuracy in diagnosis. 

\section{Conflict of Interest}
The authors declare that there is no conflict of interest that could be perceived as prejudicing the impartiality of the research reported.

\section{Author Contributions and IRB Information}

E.X. designed and conducted the experiments and analyzed the data. E.X., T.W. drafted the manuscript.  T.W., V.P. assisted with data analysis. V.P. assisted with model implementation. E.X. conducted human subject trials. All authors have read and agreed to the published version of the manuscript.  

The data that supports the findings of this study is available from the corresponding author, E.X., upon reasonable request and approval by the ethics committee.

The IRB/Ethics Committee of Shady Side Academy gave ethical approval for this work. The IRB consisted of an administrator, teacher, RN, MD, psychologist, and research professional. No members of the IRB were personally related to any author. All participants signed a consent form developed by the authors in conjunction with the IRB and Designated Supervisor.

\section{Supplementary Figures}
All Bland-Altman plots are found in Supplementary Figures 1-12. The Bland–Altman plots show the agreement between the measured values from the Apple Watch versus the Commercial ECG in terms of Duration(ms) and Amplitude (mV). The horizontal dotted red line in the plots shows the mean of the differences (=bias) between the two methods. The green horizontal dotted lines show the upper and lower 95\% limits of agreement (= bias ± 1.96 × SD). 

\newpage
\bibliographystyle{unsrtnat}
\bibliography{ecg}

\end{document}